# Phosphorus Pentamers: Floating Nanoflowers form a 2D Network

*Wei Zhang, Hanna Enriquez, Yongfeng Tong, Andrew J. Mayne, Azzedine Bendounan, Yannick J. Dappe, Abdelkader Kara, Gérald Dujardin, and Hamid Oughaddou\**

W. Zhang, Dr. H. Enriquez, Dr. A. J. Mayne, Dr. G. Dujardin, Prof. H. Oughaddou
Université Paris-Saclay, CNRS, Institut des Sciences Moléculaires d'Orsay, 91405 Orsay, France
\*E-mail: Hamid.oughaddou@u-psud.fr

Dr. Y. F. Tong, Dr. A. Bendounan
TEMPO Beamline, Synchrotron SOLEIL, L'Orme des Merisiers Saint-Aubin, B.P.48, F-91192 Gif-sur-Yvette Cedex, France

Dr. Y. J. Dappe
Université Paris-Saclay, CNRS, CEA, Service de Physique de l'Etat Condensé, 91191 Gif-sur-Yvette, France

Prof. A. Kara
Department of Physics, University of Central Florida, Orlando, FL 32816, USA

Prof. H. Oughaddou
Département de physique, Université de Cergy-Pontoise, F-95031 Cergy-Pontoise Cedex, France

**Abstract:**

We present an experimental investigation of a new polymorphic 2D single layer of phosphorus on Ag(111). The atomically-resolved scanning tunneling microscopy (STM) images show a new 2D material composed of freely-floating phosphorus pentamers organized into a 2D layer, where the pentamers are aligned in close-packed rows. The scanning tunneling spectroscopy (STS) measurements reveal a semiconducting character with a band gap of 1.20 eV. This work presents the formation at low temperature (LT) of a new polymorphic 2D phosphorus layer composed of a floating 2D pentamer structure. The smooth curved terrace edges and a lack of any clear crystallographic orientation with respect to the Ag(111) substrate at room temperature indicates a smooth potential energy surface that is reminiscent of a liquid-like growth phase. This is confirmed by density functional theory (DFT) calculations that find a small energy barrier of only 0.17 eV to surface diffusion of the pentamers (see Supplemental Material). The formation of extended, homogeneous domains is a key ingredient to opening a new avenue to integrate this new 2D material into electronic devices.

# 1. Introduction

The great success of graphene has inspired an explosive increase in the exploration of novel two-dimensional (2D) materials.[1-4] Examples include silicene,[5-8] transition-metal dichalcogenides,[9-11] germanene,[12-14] borophene[15,16], stanene[17], and phosphorene.[18,19] Among them, phosphorene, made of one or few atomic thick phosphorus layers, has received particular attention recently due to its exotic properties of high carrier mobilities,[19,20] inherent tunable direct band gap (from bulk ~0.3 eV to monolayer ~1.8 eV),[21-26] high on/off ratios,[19,26] and high in-plane anisotropy.[27,29]

Black phosphorus as the most stable allotrope of phosphorus,[30] is composed of layers stacked together by weak van der Waals forces.[31] The exfoliation method from a 3D crystal can be used to produce a few layers of black phosphorene.[18,19,23] While each phosphorus atom is covalently bonded to three adjacent atoms, these phosphorene layers prefer to form a corrugated sheet structure unlike the planar graphene.[32]

The molecular beam epitaxy (MBE) technique has proven to be advantageous in producing phosphorene on metal substrates. Blue phosphorene can be grown on Au(111),[33-38] while only phosphorus clusters are reported on Cu(111).[39,40] Thus, the effect of the substrate is an important factor determining the nature of the phosphorene growth. A moderate interaction strength has been found for a P layer adsorbed on the Ag(111) surface.[41,42] However, until now the actual production of epitaxial phosphorus sheets on Ag(111) has not yet been reported. Theoretical calculations predict that phosphorene can form many polymorphic structures that do not necessarily have a hexagonal structure.[43] Indeed, earlier calculations indicate that pentameric structures are stable.[44]

In this study, for the first time, we report the epitaxial growth of single layer phosphorus on Ag(111) surface under ultrahigh-vacuum (UHV) conditions. Our results demonstrate that the phosphorus is organized in the form of pentamers presenting a band gap of 1.2 eV. This is the first observation of such a structure made by phosphorus. The detailed morphology is characterized by *in situ* scanning tunneling microscopy (STM) operated at 78 K. The chemical composition is determined by *in situ* Auger electron spectroscopy (AES) and X-ray photoelectron spectroscopy (XPS). The electronic structure of the phosphorus pentamers on Ag(111) is probed by scanning tunneling spectroscopy (STS). The adsorption and diffusion of P-pentamers on Ag(111) were calculated using DFT with self-consistent van der Waals interactions.

## 2. Results and Discussion

*In situ* AES was carried out during the deposition process. Typical AES spectra of Ag(111) surface before and after 0.5 monolayer (ML) of phosphorus deposition are presented in Figure **1**. The characteristic spectrum of Ag MNN acquired on clean Ag(111) presents sharp peaks at 351 and 356 eV and broadened secondary peaks at 260 and 302 eV.[45] After deposition of phosphorus, the P LMM peak is clearly observed at the expected energy position of 120 eV. Moreover, none of the features show any characteristic shifts or satellite peaks that would appear if surface alloying or compound formation had occurred.

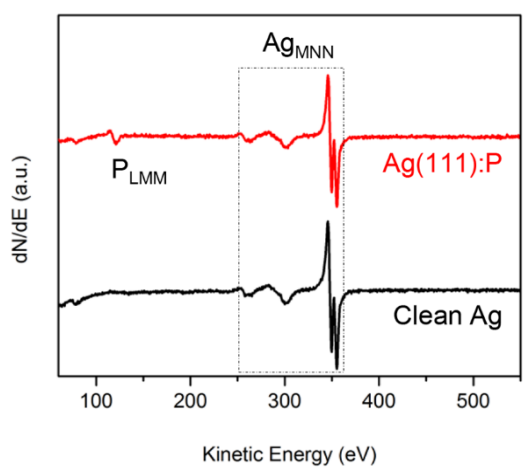

**Figure 1.** AES spectra of Ag(111). Spectra collected from clean (black line) and after 0.5 ML of phosphorus deposition (red line).

The LEED pattern recorded at room temperature (RT) did not show any extra diffraction spots related to the phosphorus deposition (Figure S1a-b, Supporting Information). This suggests that the phosphorus layer is disordered. STM experiments performed at RT to investigate the structure of the phosphorus layer reveal a flat 2D phosphorus layer in the large-scale STM images (Figure S1c, Supporting Information). The smooth round edges to the layers have no clear crystallographic orientation with respect to the Ag(111) substrate reminiscent of a liquid-like growth phase. This liquid behavior has been observed in the growth of Perlyene on Ag(110).[46] Furthermore, a close-up STM image of the phosphorus layer clearly shows no order even at the nanometer scale (Figure S1d, Supporting Information). The RT-STM experiments confirm the initial analysis of the LEED pattern.

An investigation of the growth of the Phosphorus layer was made at low temperature. After P deposition, the sample was cooled down to 78 K and analyzed in the STM. A typical large-area STM image is shown in **Figure 2**a. In comparison with the image recorded at RT, the phosphorus appears to have undergone a phase change from a liquid-like layer to a crystalline structure. The phosphorus deposition results in a single layer of pentamers that are one atomic layer thick and forms different rotational domains covering the terraces (highlighted with different colors). Comparison of the orientation of the pentamer domains (Figure 2a) relative to the metal substrate (inset in Figure 2a) yields an angle of 19°. All distinct domains have sizes in the 50-300 nm range, exhibit a striped structure rotated by 19° relative to the three main crystallographic directions of the Ag(111) substrate, leading to six different domains. At RT, we observe that thermal diffusion of the pentamers is facile, owing to a smooth potential energy surface with shallow binding sites. As the temperature is lowered, self-assembly becomes energetically favorable so that an ordered layer is observed in the STM images. This competition between thermal diffusion and ordering occurs at the molecular level, for example in benzene on silicon.[47] All these observations point to a weakly corrugated potential energy surface between the phosphorus layer and the Ag(111) substrate. The atomically resolved STM image (Figure 2b) shows that the phosphorus sheet is composed of a series of rows. Careful inspection of the STM image reveals that each small structure is composed of 5 atoms in the form of a pentamer. The average distance between the nearest-neighbor pentamers is around 0.76 nm from the line profile in Figure 2c (corresponding to the blue line in Figure 2b). This value is √7 × the unit cell of Ag(111) (√7 × 0.289 nm = 0.765 nm). In addition, Figure 2b shows that pentamers are close-packed into rows. The pentamers within the same row have the same orientation while a neighboring row has a different orientation giving a periodic pattern to the stripes (see **Figure 3** for a detailed description). The line profile measured across these stripes (red line in Figure 2b) gives a periodicity of 4.53 nm which is very close to 6√7 times the unit cell of Ag(111) (6√7 × 0.289 = 4.6 nm). Hence The phosphorus pentamer structure presents a (√7 × 6√7)R19° unit cell with respect to Ag(111). The unit cell is indicated by black parallelogram.

The periodicity relationship is confirmed in the fast Fourier transform (FFT) of the STM images. Figure 2e presents the FFT pattern of the clean Ag(111) surface, where the six symmetric bright spots related to the pristine Ag(111) substrate are highlighted in white circles. Figure 2f shows the FFT pattern of Figure 2b and gives only the periodicity of the pentamers structure as only pentamers are shown in the STM image. The spots can be assigned to the phosphorus superstructure; the green and yellow circles highlight diffraction spots corresponding to √7 and 6√7, respectively. The FFT of the two STM images gives independent confirmation as the rotation of 19° is clearly visible in reciprocal space (see Figures 2e and 2f) in agreement with the analysis of the STM images (see Figure 2a).

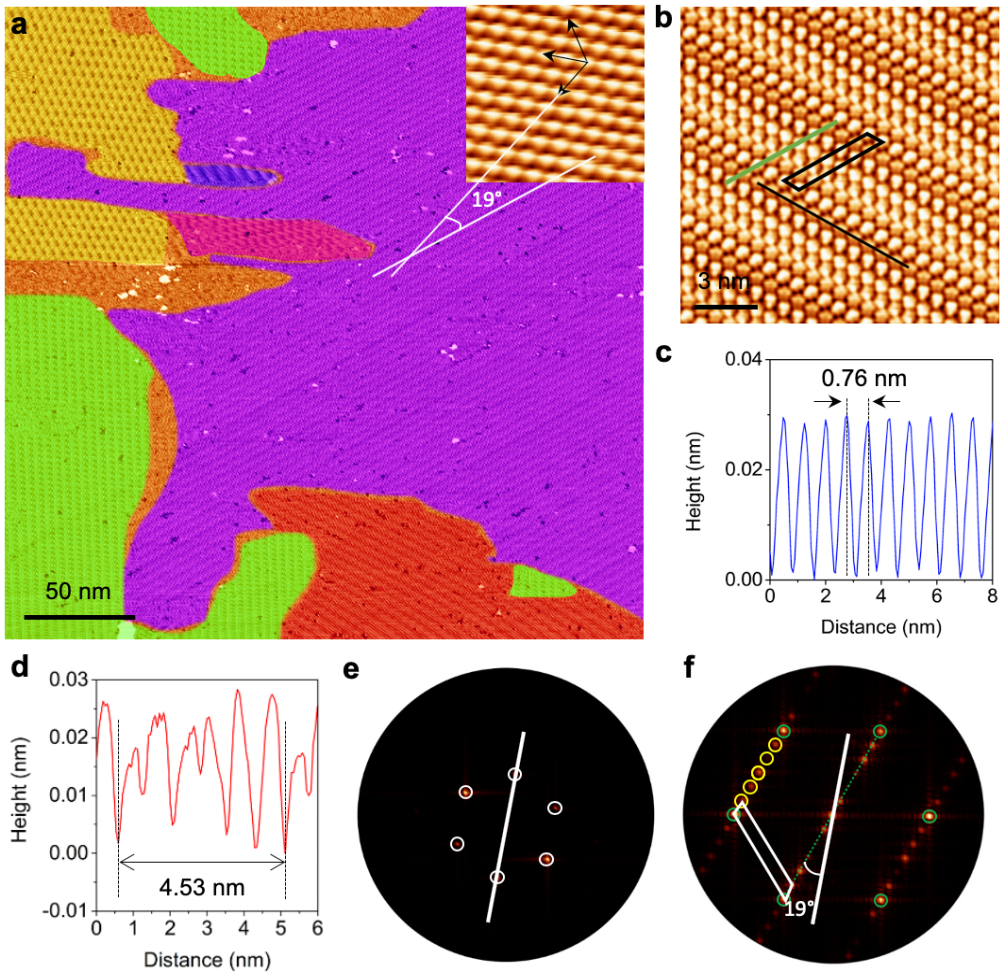

**Figure 2.** Formation of 2D phosphorus sheets on Ag(111). a) Topographic STM overview map revealing that the phosphorus covers the Ag(111) terrace and forms different domains highlighted by different colors (U = −200 mV, I = 0.80 nA). The inset shows an atomically resolved STM image of clean Ag(111) surface, the black arrows indicate [1−10] crystal directions of the Ag lattice. b) Zoom of a showing a striped-phase atomic-scale structure of the phosphorus adlayer (U = −500 mV, I = 1.00 nA). The unit cell is indicated in black. c) Line profile taken along the black line in (a), revealing the distance between nearest-neighbor pentamers of 0.76 nm. d) Line profile measured across the stripes (green line in image (b)), revealing the periodicity of about 4.53 nm. e) Fast Fourier transform (FFT) of clean Ag(111). The white circles represent the (1 × 1) Ag(111) lattice. f) FFT of (b). The green and yellow circles correspond to × √7 in one direction and × 6√7 in the other direction relative to Ag. The rotation between the unit cell of silver and the pentamer superstructure (19°) is indicated. The unit cell in the reciprocal space is also indicated in white.

The high-resolution STM image in Figure 3a reveals the finer details of the pentamer structure of phosphorus on Ag(111), where different orientations of pentamers are observed. One of the unit cells is indicated by the black parallelogram. Along the direction **a**, the pentamers are similar and have the same relative alignment. However, along the direction **b**, each pentamer is rotated by 36° relative to its neighbor as highlighted by the red and blue pentagons in Figure 3a. The line profile (Figure 3b) measured along the lines A and B in Figure 3a gives a buckling of the pentamers in the range 0.09 - 0.23 Å (very close to the calculated buckling of about 0.3 Å, see Supplemental Material), even the STM technique is sensitive to both topology and electron density. The average distance measured between the P atoms within each pentamer is around 0.27 nm. This value is larger than the 0.224 nm reported for the phosphorene sheet.[20] Furthermore, the five protrusions of each pentamer show an asymmetry in height; three bright and two dark indicating a clear buckling of the P atoms in each pentamer.

STS measurements were performed to investigate the electronic structure of this new structure. The differential conductivity (dI/dV) spectrum recorded on the phosphorus pentamers is shown in Figure 3c using the lock-in technique collected at random positions on top of pentamers. The representative dI/dV spectrum shows a semiconductor character of the phosphorus structure. In the spectrum the peaks are located at −2.0 eV and −1.0 eV below the Fermi level. The dI/dV spectrum reveals a band gap of ~1.2 eV for the new phase of phosphorus pentamers. The finite intensity observed between -0.5 and -1 V is due to the contribution of the silver substrate beneath pentamers to the total density of states probed by the STM tip. Because the metal states are closer to the Fermi level, they leak into the band gap of the phosphorus pentamer layer. This has been observed in thin insulating films on metals. [48,49]

As already reported for black phosphorene the band gap is related to the thickness of the layer.[26] The band gap value of the pentamer structure is close to the one obtained experimentally for blue phosphorene,[33,35,37] but smaller than the one obtained for an exfoliated black phosphorene monolayer (2.0 eV).[25] However, like black phosphorene, the phosphorus pentamers show a high in-plane anisotropy as the five protrusions within the pentamer show a height asymmetry.

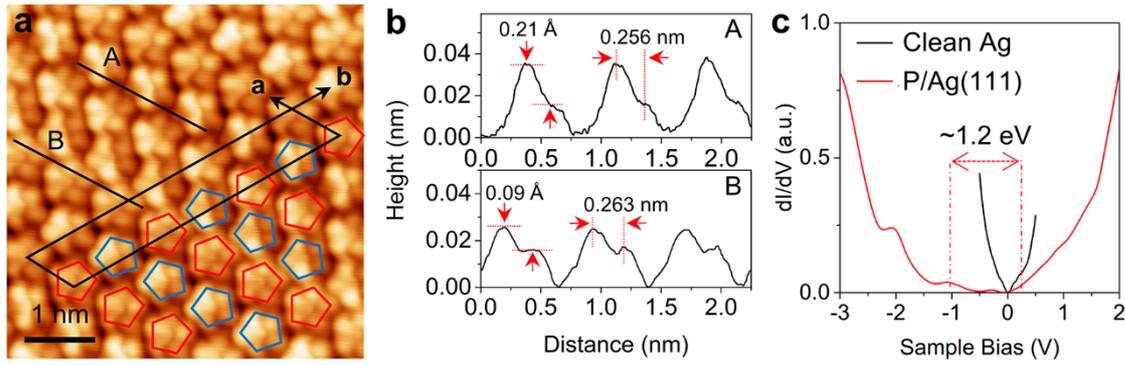

**Figure 3.** Structure of phosphorus pentamers. a) Atomically resolved filled state STM imageshowing buckled pentagonal lattice structure (U = −100 mV, I = 3.0 nA). b) Line profile along black lines in (a), revealing a corrugation in the phosphorus pentamers. c) STS dI/dV spectrum acquired on phosphorene sheet, reveals a semiconductor characteristic with a band gap of 1.2 eV. The dI/dV curve for the clean Ag(111) surface is included for comparison.

The temporary formation of pentamer vacancies is observed in the STM images. **Figure 4**a-d depicts a series of STM images taken on the same area of the phosphorus layer. Some vacancies having the form of a single missing phosphorus pentamer remain stationary during the tip scan (highlighted by the red circle). The fact that vacancy has the size and shape of a pentamer indicates that the pentamer is the fundamental building block of this structure. Vacancies are observed to undergo diffusion within the layer (marked by the blue dashed circle). This diffusion process points to two interesting properties of the pentamers: the pentamer entity is stable on its own, and the diffusion is possible even at 78 K. The presence of vacancies and their diffusion even at low temperature shows that the pentamer-silver potential energy surface is smooth with very weak corrugation because vacancies can appear and disappear within the layer. At RT, thermal diffusion dominates the STM images because not only the structure is disordered and liquid-like but the atomic resolution of the pentamers cannot be obtained. The long-range order is determined by the interaction between pentamers. There is no evidence of individual pentamers on top of the P-layer, indicating that diffusion is confined to within the layer. This has been observed in the case of atomic vacancies,[50] and molecular vacancies in a molecular layer.[51,52] This is in stark contrast to metal surfaces where diffusion proceeds via adatoms on the surface.[53]

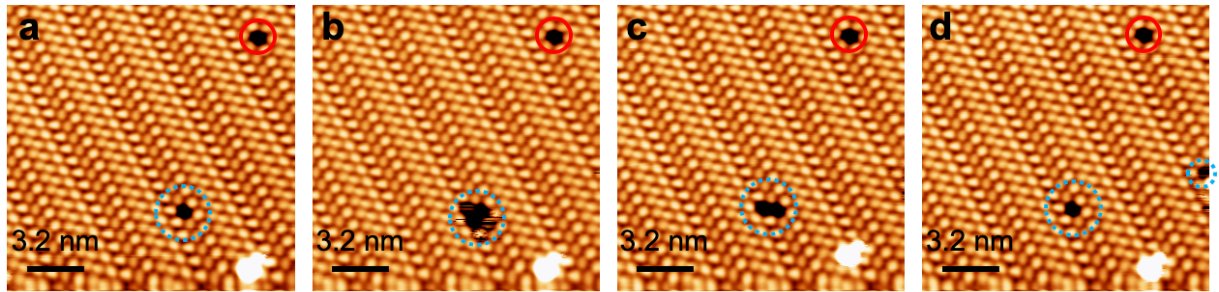

**Figure 4.** a-d) Consecutive STM images of the same area of the single layer of phosphorus on Ag(111) surface. One vacancy (red circle) remains at the same position in all images, others (blue circle) diffuse on the surface (U = −700 mV, I = 1.10 nA).

Further analysis of the nature of the chemical interaction between the phosphorus and Ag(111) surface, was obtained from XPS measurements. All spectra are fitted with a Doniach-Sunjic line shape.[54] The experimental data are displayed with black dots, while the fitted curves are in black solid lines. **Figure 5**a-b shows the characteristic XPS Ag 3d core-level spectra acquired on the clean Ag(111) surface and after formation of phosphorus pentamers. For the clean substrate, the characteristic signals of Ag $3d_{3/2}$ and $3d_{5/2}$ are located at binding energies of 374.2 eV and 368.2 eV, respectively. The best fit was obtained with only one component (red curve) with a 140 meV Gaussian profile and a 280 meV Lorentzian profile, while the spin-orbit splitting is 6 eV.

After phosphorus deposition, the shapes and energy positions of these two representative peaks remain unchanged. We also observed no splitting of these Ag 3d peaks, indicating the weak chemical interaction between the phosphorus and the underlying Ag substrate. The XPS P 2p core-level spectra obtained from phosphorene recorded at normal and grazing emission are respectively shown in Figure 5c, 5d. The best fit was obtained with a 200 meV Gaussian profile and a 80 meV Lorentzian profile, while the spin-orbit splitting is 0.865 eV. The P 2p core-level spectrum in Figure 5c is composed of two components S1 and S2 located respectively at 129.72 and 129.04 eV. These two components are assigned to two types of phosphorus atoms with different chemical environments. This can be explained by the buckled structure of phosphorus pentamers observed by STM. One component corresponds to P atoms in contact with the substrate and the second corresponds to P atoms in contact only with P atoms. The buckling in the pentamer can be explained by the strong physisorption (van der Waals interaction) of the pentamers on Ag(111), where the calculated binding energy was found to be 0.78 eV/atom (see Supplemental Material).

With increasing emission angle, the XPS surface sensitivity is enhanced to selectively detect the surface and sub-surface contributions. The relative intensities of the surface component peak will be increased. Thus, from the Figure 5d we can conclude that the S1 component is assigned to the phosphorus atoms buckled above the plane and the S2 component is assigned to the phosphorus atoms buckled below the plane. These results confirm the buckled structure observed in the STM images. In addition, from the XPS data we can estimate the ratio of the area of the two components S1 and S2 of the P 2p core level. In normal emission, the ratio S1/S2 is equal to 0.64 while for the grazing emission it is equal to 1.36. These values are consistent with 3 P atoms up and 2 P down in each pentamer, in agreement with the STM images showing three bright and two dark protrusions.

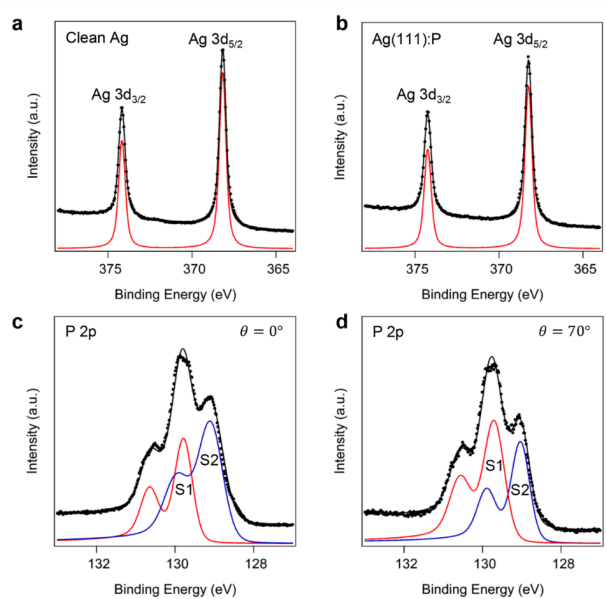

**Figure 5.** XPS analysis for the single layer of phosphorus on Ag(111) substrate. a) The Ag 3d core-level spectra of clean Ag(111) surface. b) After formation of phosphorus pentamers on Ag(111). In both spectra, the best fit is obtained with only one component (red curves). c,d)

The P 2p spectra taken at (c) normal emission and (d) grazing emission (70° off-normal), showing two asymmetric components (S1) and (S2). For all spectra, the dots correspond to data and black line overlapping the data corresponds to the best fit.

3. Conclusion

We have synthesized successfully a high-quality single layer of phosphorus on the Ag(111) substrate using MBE; this new 2D phase is composed of phosphorus pentamers. STM measurements demonstrated a characteristic array of buckled pentamer structure of phosphorus atoms on Ag(111) with anisotropic corrugation. STS investigations determined the semiconductor character of phosphorus pentamers with a band gap of 1.20 eV close to that of silicon, making this material interesting for electronic and opto-electronic devices. XPS measurements and AES analysis reveal a weak chemical interaction between phosphorus and Ag(111). However, the buckling in the pentamer points to a strong physisorption that is confirmed by DFT+vdW calculations. Moreover, XPS further confirmed the buckled conformation of phosphorene. This work reveals a new 2D polymorphic structure of phosphorus, opening new avenues for this novel 2D material.

4. Experimental Section

The experiments were performed in a commercial ultra-high vacuum (UHV) system with a base pressure of below $1.0 \times 10^{-10}$ mbar for surface preparation and characterization. The single-crystal Ag(111) substrate (99.999%) was cleaned by several cycles of Ar$^+$ sputtering (650 eV, $1 \times 10^{-5}$ mbar) followed by annealing at the temperature of 500°C for 40 minutes. During growth process, a Knudsen source loaded with black phosphorus was used to deposit P atoms on Ag(111) surface, maintained the substrate at 150°C by radiative heating from a filament located behind the sample. STM measurements were conducted in an Omicron Scanning Tunneling Microscope, operating at 78 K or 300 K. All STM experiments were acquired in constant current mode, using electrochemically etched W tips for surface characterization. STS experiments were acquired using the lock-in technique, with an applied modulation voltage of 10 mV at a frequency of 5127.7 Hz. The photoemission measurements were performed at the TEMPO beam-line of Synchrotron SOLEIL-France using a Scienta SES-2002 electron spectrometer at RT with energy resolution better than 50 meV.[55] The calibration of the XPS measurements was made with the Ag 3d core level corresponding to the clean Ag substrate. The photon energy of XPS experiments were recorded at 470 eV for Ag 3d and 280 eV for P 2p, respectively.

## 5. Computational Section

To study the adsorption of P-pentamers on Ag(111) we used the Vienna Ab initio Simulation Package (VASP) version 5.4.4.[56-58], which contains the projector augmented wave (PAW) method[59-60]. To model the exchange-correlation interaction we employed the optB88-vdW functional[61]. This particular functional includes a nonlocal implementation of the correlation energy term with the aim of seamlessly and self-consistently incorporating the long-range van der Waals (vdW) interaction. In the past, several studies demonstrated the importance of vdW interactions (including the use of optB88) in a myriad of systems[62-65], including organic molecules on transition metal substrates. To model the gold surface, we created a 4-layer slab with each layer containing 12 atoms. VASP uses periodic boundary conditions so we placed at least 20 Å of vacuum in between slabs. A lattice constant of 4.147 Å was taken. To achieve structural relaxation, we utilized the conjugate gradient method. The force criterion was set to 0.02 eV/Å, the plane wave energy cutoff was chosen at 400 eV, and the Brillouin zone was sampled using a 4x4x1 k-points mesh. To simulate the diffusion, we used the Nudged Elastic Band method, with Climbing Image (NEB+CI), using 5 intermediary images[66].

**Supporting Information**

Supporting Information is available from the Wiley Online Library or from the author.

**Acknowledgements**

W.Z. would like to thank the China Scholarship Council (CSC) for the PhD financial support (scholarship). The computational work of AK is supported by the U.S. Department of Energy